\documentclass[11pt,a4paper]{article}

\usepackage[T1]{fontenc}
\usepackage[utf8]{inputenc} 
\usepackage{graphicx}
\usepackage{amsmath,amssymb}
\usepackage{cite}
\usepackage[colorlinks=true,linkcolor=blue,citecolor=blue,urlcolor=blue]{hyperref}
\usepackage{orcidlink}
\usepackage{QCDEX2026}

\newcommand{\bea}{\begin{eqnarray}}
\newcommand{\eea}{\end{eqnarray}}
\newcommand{\be}{\begin{equation}}
\newcommand{\ee}{\end{equation}}

\newcommand{\ar}{a_s}

\begin{document}

\noindent\includegraphics[width=0.4\textwidth]{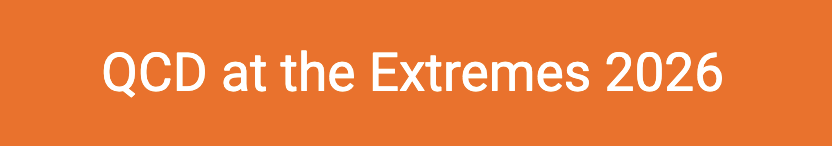}\par
\vspace{0.6em}
\begin{center}
 {\LARGE\bfseries On Gross-Llewellyn Smith sum rule with analytic coupling\par}
 \vspace{0.8em}
 {\large
I.A.~Zemlyakov$^{1}$\orcidlink{0000-0003-3887-5518},\quad I.R.~ Gabdrakhmanov$^{2}$\orcidlink{0009-0009-9372-4721},\quad N.A.~Gramotkov$^{2,3}$\orcidlink{0009-0005-9333-808X},\quad A.V.~Kotikov$^{2,*}$\orcidlink{0000-0002-1408-2735},
\quad  O.V.~Teryaev$^{2}$\orcidlink{0000-0001-7002-9093}  \quad and\quad D.A.~Volkova$^{2,4}$\orcidlink{0009-0001-6262-7154}

 \par}
 \vspace{0.5em}
 {\small
$^{1}$Tomsk State University, 634010 Tomsk, Russia\par
 $^{2}$Joint Institute for Nuclear Research, 141980 Dubna, Moscow region, Russia\par
$^3$Moscow State University, 119991, Moscow, Russia\par
 $^4$Dubna State University,
 141980 Dubna, Moscow Region, Russia\par}
 
\vspace{0.35em}
 {\small $^{*}$Corresponding author: \href{mailto:kotikov@theor.jinr.ru}{kotikov@theor.jinr.ru}\par}
 \vspace{0.35em}
 {\small\itshape Contribution to the proceedings of the International Workshop
 	"QCD at the Extremes 2026" (QCDEX2026)\par}
\end{center}
\vspace{0.5em}

\begin{abstract}
We present the main results of our recent paper \cite{Gabdrakhmanov:2025vxw}, in which we investigated the Gross--Llewellyn Smith sum rule within the framework of analytic QCD.
That study compared experimental data and lattice calculations with predictions from perturbative QCD, using both conventional and analytic approaches to perturbation theory.
We found that conventional perturbation theory fails to reproduce the data, whereas the analytic approach yields good agreement with
experiment and lattice calculations. We also discussed the relationship between the Gross--Llewellyn Smith and Bjorken sum rules.

\end{abstract}


\vspace{1.0cm}

\section{Gross-Llewellyn Smith sum rule}

The Gross--Llewellyn Smith (GLS) sum rule is defined by the integral \cite{Gross:1969jf} 
\be
C_{GLS}^{p+n}(Q^2)=\frac{1}{2}\, \int\limits^1_0\left[F_3^{\overline{\nu}p}(x,Q^2)+F_3^{\nu n}(x,Q^2)\right]dx\,,
 \label{GLS.I}
 \ee
In the quark--parton model, this integral counts the number of valence quarks in a proton. 
The structure function $F_3$ contributes to deep-inelastic neutrino scattering and plays an important role in our understanding of both QCD and electroweak processes.

The GLS sum rule can be expressed theoretically as follows:
\be
C_{GLS}^{p+n}(Q^2)=
3 \, \bigl(1-D(Q^2)\bigr) + \frac{\hat{\mu}_{4} M^2}{Q^{2}+M^2}\,.
\label{Gpn.mOPEa}
\ee
Here, $(1-D(Q^2))$ is the leading-twist (twist-two) contribution, while the coefficients $\hat{\mu}_{4}$ and $M^2$ of
the so-called "massive" twist-four term (see \cite{Teryaev:2013qba,Khandramai:2016kbh} are free parameters to be determined from experimental data.
We use the ``massive'' twist-four term because we focus on relatively low values of $Q^2$.

Up to the $k$th order of perturbation theory (PT) 
(hereafter we use $\overline{MS}$-scheme),
the twist-two part has the form
\bea
D^{(1)}(Q^2)=\frac{4}{\beta_0} \, a^{(1)}_s,~~D^{(k\geq2)}(Q^2)=\frac{4}{\beta_0} \, a^{(k)}_s\left(1+\sum_{m=1}^{k-1} d_m \bigl(a^{(k)}_s\bigr)^m
\right)\,,
\label{GLS} 
\eea
where
\be
a^{(k)}_s(Q^2)=\frac{\beta_0 \alpha^{(k)}_s(Q^2)}{4\pi}=\beta_0\,\overline{a}^{(k)}_s(Q^2)
\label{ak}
\ee
and
$d_1$,
$d_2$,
and $d_3$
are known exactly (see \cite{Baikov:2010je} and references therein).
Hereafter, $\beta_0$ is the first coefficient of the QCD $\beta$-function:
\be
\beta(\overline{a}^{(k)}_s)\equiv \frac{d \overline{a}^{(k)}_s(Q^2)}{dL}
=-{\left(\overline{a}^{(k)}_s\right)}^2 \bigl(\beta_0 + \sum_{i=1}^k \beta_i {\left(\overline{a}^{(k)}_s\right)}^i\bigr),
\label{bQCD}
\ee
where $\beta_i$ are known up to $k=4$ \cite{Baikov:2008jh}.

Following \cite{Cvetic:2006mk}, we introduce the derivatives (at the $k$-th order of PT)
\be
\tilde{a}^{(k)}_{n+1}(Q^2)=\frac{(-1)^n}{n!} \, \frac{d^n a^{(k)}_s(Q^2)}{(dL)^n},~~L=\ln \frac{Q^2}{\Lambda^2}\,,
\label{tan+1}
\ee
which play a key role in the construction of analytic QCD (see, e.g., \cite{Kotikov:2022swl}). 

Instead of expanding in powers of $\ar$, one can use a series of derivatives, $\tilde{a}_{n}(Q^2)$. 
Indeed, although each differentiation lowers the power of $\ar$, it also introduces an additional factor of the $\beta$ function and, consequently, an additional factor of $\ar^2$.
By definition (\ref{tan+1}), the expressions for $\tilde{a}_{n}$ and $\ar^n$ coincide at leading order (LO).
Beyond LO, a one-to-one correspondence between $\tilde{a}_{n}$ and $\ar^n$ was established in 
\cite{Cvetic:2006mk, Cvetic:2010di} and
extended to the fractional case in \cite{Cvetic:2011ym}.

Rewriting the powers of the coupling constant
in terms of its derivatives, we obtain
\bea
D^{(1)}(Q^2)=\frac{4}{\beta_0} \, \tilde{a}^{(1)}_s,~~D^{(k\geq2)}(Q^2)=\frac{4}{\beta_0} \, \tilde{a}^{(k)}_s\left(1+\sum_{m=1}^{k-1} \tilde{d}_m \bigl(\tilde{a}^{(k)}_s\bigr)^m
\right)\,,
\label{GLS.1}
\eea
where
\be
\tilde{d}_1=d_1,~~\tilde{d}_2=d_2-b_1d_1,~~\tilde{d}_3=d_3-\frac{5}{2}b_1d_2-\bigl(b_2-\frac{5}{2}b^2_1\bigr)\,d_1
\label{tdi} 
\ee
and $b_i=\beta_i/\beta_0^{i+1}$.

For three active quark flavors ($f=3$), as assumed in this work, the coefficients are
\bea
d_1=\tilde{d}_1=1.59,~~d_2=3.75~(\tilde{d}_2=2.51),~~d_3=16.77~(\tilde{d}_3=10.44)\,,
\label{td123} 
\eea
Thus, the coefficients in the derivative expansion are slightly smaller.

Using the derivatives of the coupling constant defined in Eq. (\ref{tan+1}), we can write the results in Eq. (\ref{Gpn.mOPEa}) in analytic perturbation theory (APT) as follows:
\be
C_{\rm{A},GLS}^{p+n}(Q^2)=
3 \, \bigl(1-D_{\rm A}(Q^2)\bigr) + 
\frac{\hat{\mu}_{4,A} M^2}{Q^{2}+M^2}
\, ,
\label{Gpn.MAa}
\ee
where the perturbative contribution, $D_{\rm{A}}(Q^2)$
takes the same forms as (\ref{GLS.1}),
but the derivatives $\tilde{a}^{(k)}_{\nu}$
 are replaced by the corresponding analytic derivatives $\tilde{A}^{(k)}_{\nu}$
 (the corresponding expressions for the analytic coupling constants
 $\tilde{A}^{(k)}_{\nu}$ can be found in 
 \cite{Kotikov:2022sos,Kotikov:2023meh}):
\be
D^{(1)}_{\rm A}(Q^2)=\frac{4}{\beta_0} \, A^{(1)},~~
D^{k\geq2}_{\rm{A}}(Q^2) =\frac{4}{\beta_0} \, \Bigl(A^{(k)}
+ \sum_{m=2}^{k} \, \tilde{d}_{m-1} \, \tilde{A}^{(k)}_{\nu=m} \Bigr)\,.
\label{DBS.ma} 
\ee

It is important to emphasize that the perturbative structures of the GLS sum rule (\ref{Gpn.mOPEa}) and (\ref{Gpn.MAa})
are very similar to those of the polarized Bjorken sum rule (BSR). Both sum rules are governed by closely related twist-two contributions. Their approximate equivalence was discussed previously (see \cite{Broadhurst:1993ru}).
Moreover, the similarity of their twist-four contributions has also been suggested (see \cite{Kataev:2005hv}).


Therefore, following the discussion in \cite{Kataev:2005hv}, we can regard the BSR and GLS sum rules as having, approximately, the same $Q^2$ dependence, up to an overall normalization factor:
\be
C_{GLS}^{p+n}(Q^2)\approx \frac{18}{g_A}\,\Gamma_{1}^{p-n}(Q^2)\,
\label{GLS.app}
\ee
Accordingly, in analytic QCD, we obtain the analogous relation:
\be
C_{\rm{A},GLS}^{p+n}(Q^2)\approx \frac{18}{g_A}\,\Gamma_{\rm{A},1}^{p-n}(Q^2)\,.
\label{GLS.app.ma}
\ee

\section{Results}

Following the approach of Refs.
\cite{Pasechnik:2008th,Khandramai:2011zd,Ayala:2017uzx,Ayala:2018ulm,Kotikov:2012eq,Gabdrakhmanov:2023rjt,Gabdrakhmanov:2024bje,Gabdrakhmanov:2025afi}, which studied the BSR, we fit the experimental data for the GLS sum rule within the frameworks of conventional and analytic QCD (see Eqs. 
(\ref{Gpn.mOPEa}) 
and (\ref{Gpn.MAa}), respectively).
\footnote{A fairly similar analysis of the GLS sum rule with the usual form of the twist-four term can be found in
\cite{Milton:1998ct}.}

\begin{figure}[!htb]
\centering
\includegraphics[width=0.98\textwidth]{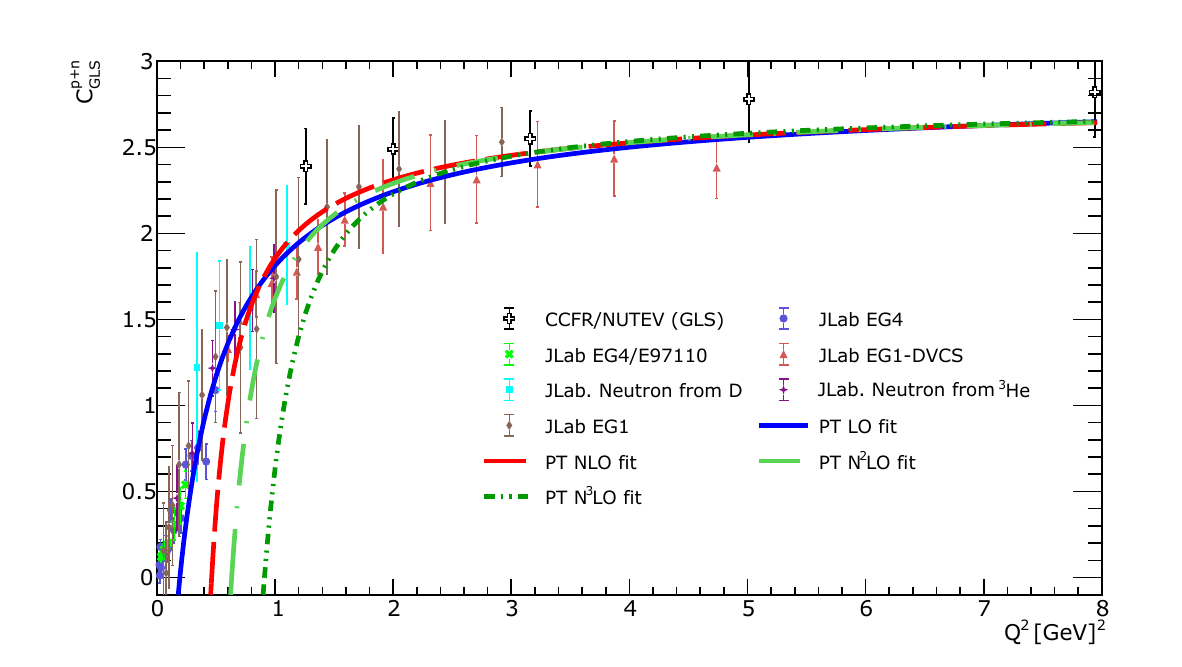}
\caption{
 \label{fig:PT2}
 The results (\ref{Gpn.mOPEa}) 
in the first three orders of PT.
}
\end{figure}

As in the BSR case (see \cite{Gabdrakhmanov:2024bje,Gabdrakhmanov:2023rjt}), for ordinary PT, 
Fig. \ref{fig:PT2} shows a discrepancy between the QCD predictions and the experimental data.
The discrepancy also increases with the order of perturbation theory, as it does in the BSR case
(see, e.g., \cite{Gabdrakhmanov:2024bje}), and the reason is the same.
As the perturbative order increases, the Landau pole of the strong coupling moves towards higher values of $Q^2$
 (see \cite{Kotikov:2022sos,Kotikov:2023meh}).
Consequently, the GLS predictions shift towards more negative values because the perturbative corrections are negative.

Because only a few experimental data points are available for the GLS sum rule, we also show experimental BSR data in Fig. \ref{fig:PT2}. These data are scaled by the factor $18/g_A$ and are not included in the fits. (see \cite{Gabdrakhmanov:2025afi} and references therein)
 scaled by the factor $18/g_A$. 

\begin{figure}[!htb]
\centering
\includegraphics[width=0.98\textwidth]{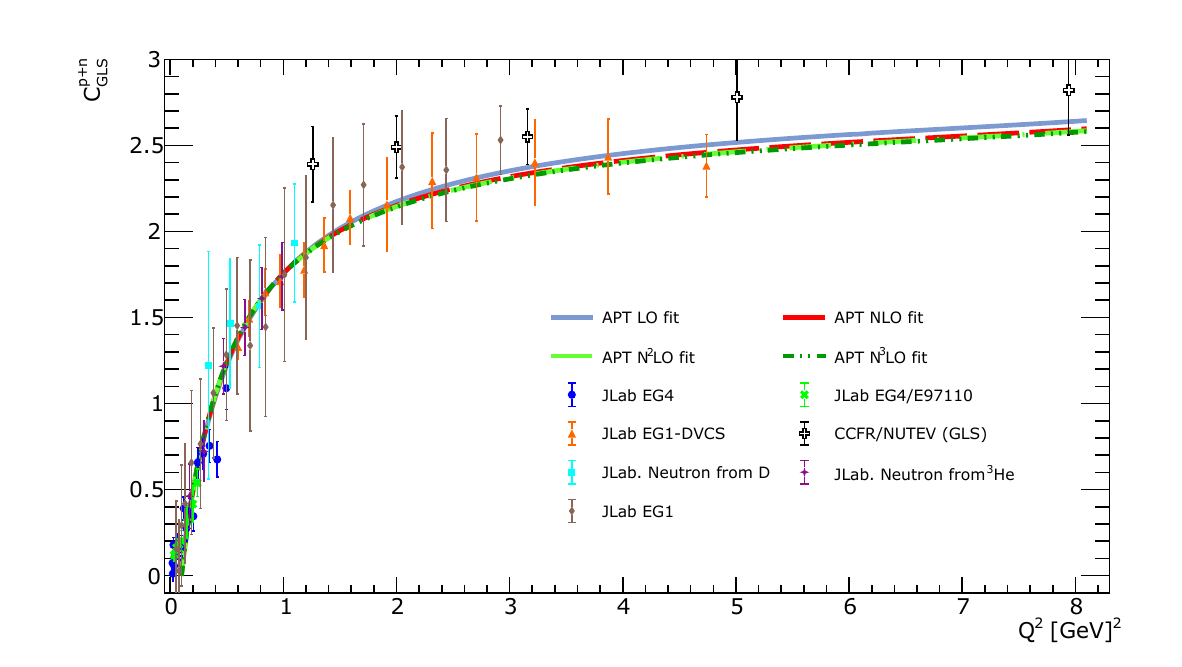}
\caption{
 \label{fig:APT1}
 As in Fig. \ref{fig:PT2} but for the APT result (\ref{Gpn.MAa})
}
\end{figure}

In analytic QCD, the theoretical predictions are in good agreement with the experimental data
(see Fig. \ref{fig:APT1}).

The available experimental information on the GLS sum rule has large uncertainties, and the number of data points is very small.
It is therefore useful to supplement this comparison with recent lattice QCD results from Ref. \cite{Can:2025zsr}.
The lattice ``data'' in \cite{Can:2025zsr} differ substantially for different values of $\beta$. In our analysis, we use the lattice data obtained at a large value of $\beta$, which corresponds to the smallest lattice spacing, $a$.

\begin{figure}[!htb]
\centering
\includegraphics[width=0.98\textwidth]{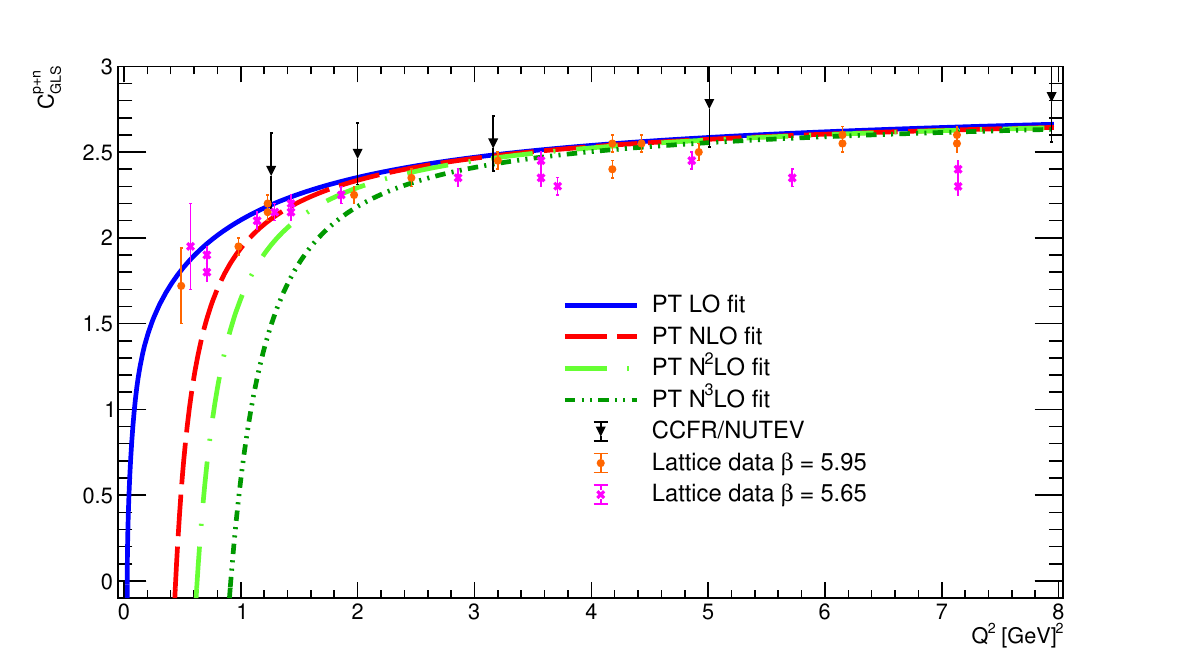}
\caption{
 \label{fig:PT1L}
 As in Fig. \ref{fig:PT2} but with lattice data.
}
\end{figure}

For conventional perturbation theory, the results of fits to these lattice data are shown in
Fig. \ref{fig:PT1L}.
The LO results are in fairly good agreement with the lattice data, whereas the results at higher perturbative orders are not.
\footnote{
In view of the close analogy with the BSR case (see \cite{Gabdrakhmanov:2024vol}), a detailed analysis of the low-$Q^2$ region may reveal that the description is inadequate even at LO. Unfortunately, no information is available for the low-$Q^2$ region of the GLS sum rule because there are no corresponding experimental or lattice data.
}
As in the comparison with experimental GLS data discussed above, the discrepancy increases with the perturbative order for the same reasons.

\begin{figure}[!htb]
\centering
\includegraphics[width=0.98\textwidth]{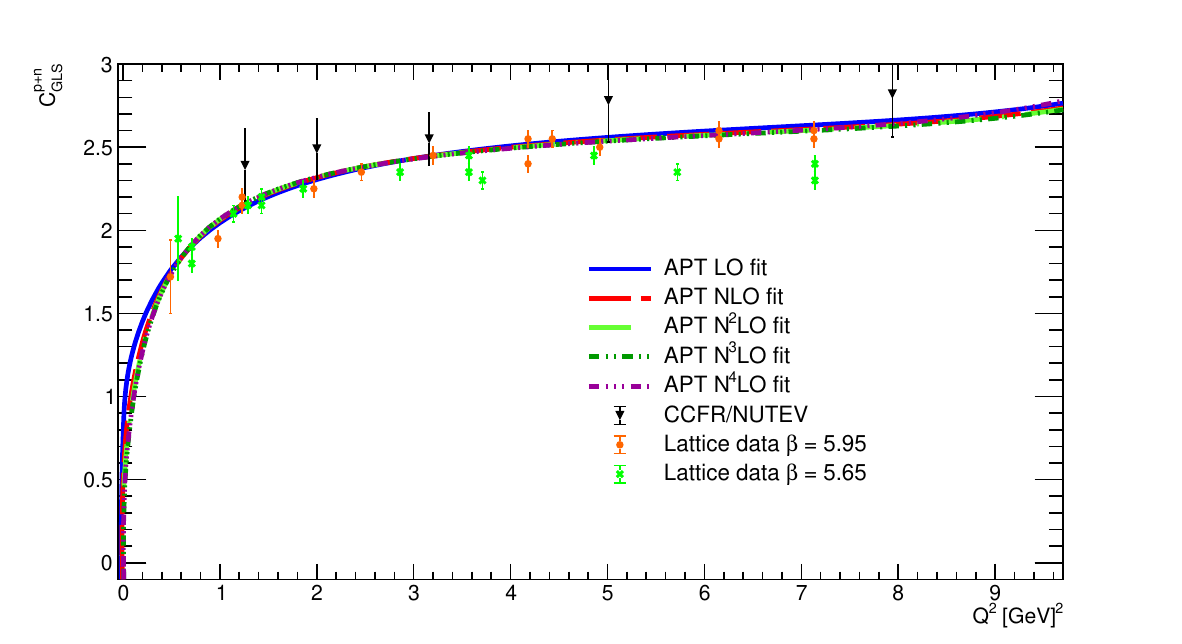}
\caption{
 \label{fig:APT2L}
 As in Fig. \ref{fig:APT1} but with the lattice ``data''.
}
\end{figure}

For analytic QCD, we find good agreement with the lattice data. The results of fits to these data are shown in Fig. \ref{fig:APT2L}.

\begin{figure}[t]
\centering
\includegraphics[width=0.98\textwidth]{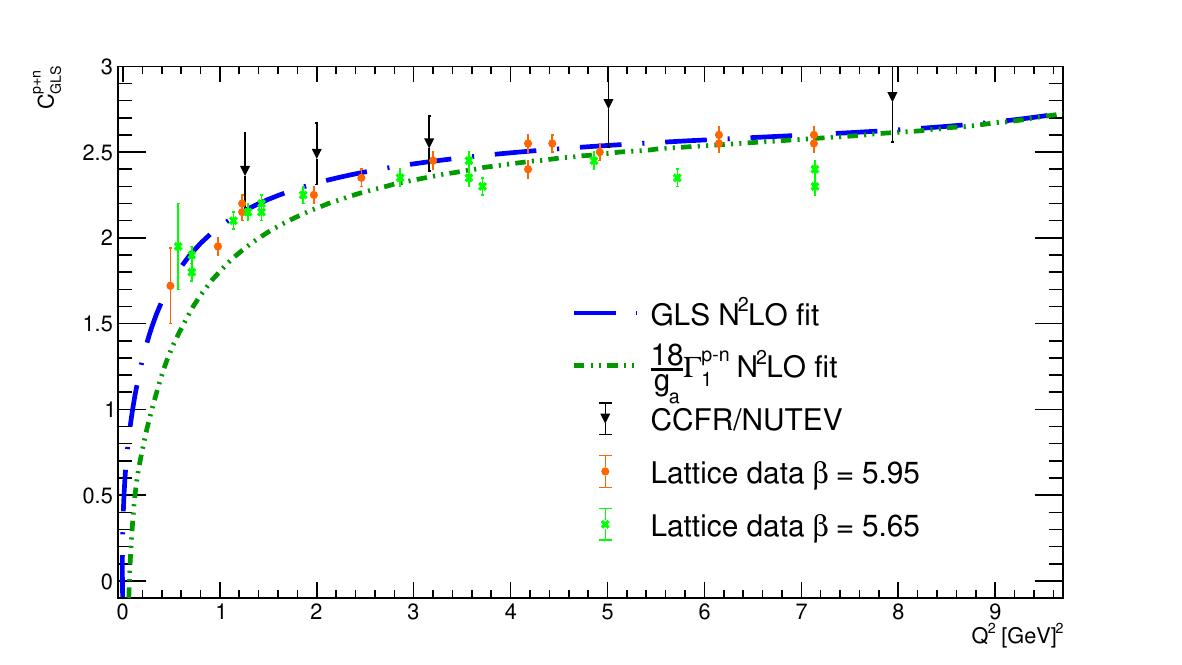}
\caption{
 \label{fig:low1}
 As in Fig. \ref{fig:APT2L}, but with
 the rescaled BSR curve (\ref{GLS.app.ma})
 (the green curve) added.
}
\end{figure}

Figure \ref{fig:low1} shows the GLS sum-rule results obtained from the lattice data, together with the BSR results scaled by a factor of $18/g_A$.
The BSR results were obtained under similar conditions in Ref. \cite{Gabdrakhmanov:2024bje}. 
As shown in Fig. \ref{fig:low1}, the relation in Eq. (\ref{GLS.app.ma}) holds only at relatively large values of $Q^2$, namely $Q^2\geq 5$ GeV$^2$. At lower values of $Q^2$, $C_{\rm{A},GLS}^{p+n}(Q^2)$ and $(18/g_A)\,\Gamma_{\rm{A},1}^{p-n}(Q^2)$ begin to differ. This difference arises from the different parameter values in the twist-four terms of the GLS and Bjorken sum rules.

\section{Conclusions}

We have presented some of the results of Ref. \cite{Gabdrakhmanov:2025vxw}, in which the Gross--Llewellyn Smith sum rule was studied within the frameworks of conventional and analytic QCD.
We compared experimental data and recent lattice calculations with perturbative QCD predictions obtained using conventional and analytic approaches to perturbation theory.
Our results show that conventional perturbation theory fails to reproduce the data beyond LO, whereas the analytic approach is in good agreement with both the experimental data and the lattice calculations.

We also discussed the relationship between the Gross--Llewellyn Smith and Bjorken sum rules.
In Ref. \cite{Gabdrakhmanov:2025vxw}, we found that the GLS sum rule is similar to the Bjorken sum rule multiplied by the factor $18/g_A$ at sufficiently large values of $Q^2$ ($Q^2\geq 5$ GeV$^2$), where the twist-two contributions dominate. At lower values of $Q^2$ ($Q^2\leq 2$ GeV$^2$), however, the two sum rules differ.

{\sl Acknowledgements.} We are grateful to the organizers of the QCD at the Extremes conference for their dedication and efforts in fostering an inclusive and closely connected scientific community.

One of us (A.V.K.) was supported in part by the Russian Science Foundation grant
No. 25-22-00576.
I.A.Z. was supported by the Fellowship ANID Beca de Doctorado Nacional No. 21250067.

\bibliographystyle{QCDEX2026}
\bibliography{zemlyakov_QCDatExtr-arxiv}
\end{document}